\documentclass[]{article}
\usepackage[a4paper, left=2cm, right=2cm, top=2cm, bottom=2cm]{geometry}
\usepackage[version=3]{mhchem}
\usepackage[utf8]{inputenc}
\usepackage{booktabs}
\usepackage{graphicx}
\usepackage{caption}
\usepackage{subcaption}
\usepackage{changepage}
\usepackage{gensymb}
\usepackage{cite}
\usepackage{datetime}
\usepackage{float}
\usepackage[british]{isodate}
\usepackage{enumerate}
\usepackage[space]{grffile}
\usepackage{setspace}
\usepackage{tabularx}
\usepackage{array,booktabs}
\pdfoutput=1

\title{Approximate solution of the time-dependent Kratzer plus screened Coulomb potential in Feinberg-Horodecki equation}
\date{}
\author{Mahmoud Farout$^{1}$, Ramazan Sever$^{2}$ and Sameer M. Ikhdair$^{1,3}$}
\begin{document}

\maketitle
$^1$ Department of Physics, An-Najah National University, Nablus, Palestine 

$^2$ Department of Physics, Middle East Technical University, 06531 Ankara, Turkey 

$^3$ Department of Electrical Engineering, Near East University, Nicosia, Northern Cyprus, Mersin 10, Turkey

\section*{\bf {\centering Abstract}}
\noindent We obtain the quantized momentum eigenvalues, $P_n$, together with space-like coherent eigenstates for the space-like counterpart of the Schrödinger equation, the Feinberg-Horodecki equation, with a combined Kratzer potential plus screened coulomb potential which is constructed by temporal counterpart of the spatial form of these potentials. The present work is illustrated with  two special cases of the general form: the time-dependent modified Kratzer potential and the time-dependent screened Coulomb potential.\\

\textbf{Keywords:} Quantized momentum states; Feinberg-Horodecki equation; the time-dependent screened Coulomb potential; and time-dependent modified Kratzer potential.

\textbf{PACS: 03.65.-w;03.65.Pm.}

\section{Introduction}
\noindent Any physical phenomenon in nature is usually characterized by solving differential equations. A good example is the time-dependent Schrödinger equation which describes quantum-mechanical phenomena, in which it dictates the dynamics of a quantum system. Solving this differential equation by means of any method results in the eigenvalues and eigenfunctions of that Schrödinger quantum system. However, solving time-dependent Schrödinger equation analytically is not straightforward except in some cases where the time-dependent potentials are constant, linear and quadratic functions of the coordinates [1-4].

The Feinberg-Horodecki (FH) equation is a space-like counterpart of the Schrödinger equation which was derived by Horodecki [5] from the relativistic Feinberg equation [6]. This equation has been demonstrated in a possibility of describing biological systems [7-8] in terms of the time-like supersymmetric quantum mechanics [9]. The space-like solutions of the Feinberg-Horodecki (FH) equation can be employed to test its relevance in different areas of science including physics, biology and medicine [7-8]. 

The space-like quantum systems with the Feinberg-Horodecki equation pay attention of many scientist especially in some branches of physics, such as in extended special relativity and in extended quantum mechanics [7, 10-12]. For example, they are used to explain the force between electric charges, the electric charge source, and the mass in a stable particle [7, 13]. Among these studies, Molski has also constructed the space-like coherent states of a time-dependent Morse potential with the Feinberg-Horodecki equation and showed that the obtained results for space-like coherent states can be used for Gompertzian systems [7].
 
Molski constructed the space-like coherent states of a time-dependent Morse oscillator on the basis of the FH quantal equation for minimizing the time-energy uncertainty relation and showed that the results are useful for interpreting the formation of the
specific growth patterns during crystallization process and biological growth [7]. In addition, Molski obtained FH equation to demonstrate a possibility of describing the biological systems in terms of the space-like quantum supersymmetry for an-harmonic oscillators [8]. Hamzavi et al [14] obtained the exact bound state solutions of the FH equation with rotating time-dependent Deng-Fan oscillator potential by means of parametric NU method. Eshghi et al [15] solved FH equation for time-dependent mass distribution (TDM) harmonic oscillator quantum system with a certain interaction applied to a mass distribution m(t) to provide a particular spectrum of stationary energies. Besides, the spectrum of harmonic oscillator potential V(t) acting on TDM m(t) oscillator was obtained. 

In the non-relativistic level, the Nikiforov-Uvarov method was used to obtain the bound state solutions of arbitrary angular momentum Schrödinger equation with the modified Kratzer potential [16]. The factorization method was also used to obtain the solution of the non-central modified Kratzer potential for the diatomic molecules [17]. The exact solutions of the Schrödinger equation with modified Kratzer and corrected Morse potentials with position-dependent mass were also found [18]. The coherent states for a particle in Kratzer type potentials are constructed by solving Feynman’s path integral [19]. Further, the exact solution of the Schrödinger equation for the modified Kratzer potential plus a ring-shaped potential was solved [20].

On the other hand, in the relativistic level, approximate solutions of the D-dimensional Klein-Gordon equation are obtained for the scalar and vector general Kratzer potential for any $l$ by using the ansatz method and the solutions of the Dirac equation with equal scalar and vector ring-shaped modified Kratzer potential were found by means of the Nikiforov-Uvarov method [21-22]. 

At the level of applications, some authors have studied the modified Morse-Kratzer potential for alkali hydrides [23], and the effect of modified Kratzer potential on the confinement of an exciton in a quantum dot [24]. Further, an analysis of the applications of the modified Kratzer potential, the bound states of two limiting cases of interest of the interactions and hence this approximation was used to obtain the solution of the Schrödinger equation for the Morse potential [25].

Very recently, a superposition of modified Kratzer potential plus screened Coulomb potential was suggested to study diatomic molecules [26].  Edet et al have obtained an approximate solution of the Schrödinger equation for the modified Kratzer potential plus screened Coulomb potential model, within the framework of Nikiforov–Uvarov method. They obtained bound state energy eigenvalues for N2, CO, NO, and CH diatomic molecules for various vibrational and rotational quantum numbers. Special cases were considered when the potential parameters were altered, resulting into modified Kratzer potential, screened Coulomb potential, and standard Coulomb potential [26]. Further, Okorie et al have solved the Schrödinger equation with the modified Kratzer plus screened Coulomb potential using the modified factorization method. They have also employed both the Greene–Aldrich approximation scheme and a suitable transformation scheme to obtain the energy eigenvalues equation and its corresponding energy eigenfunctions for CO, NO, and N2 diatomic molecules. They have used the energy eigenvalues of the modified Kratzer plus screened Coulomb potential to obtain the vibrational partition functions and other thermodynamic functions for the selected diatomic molecules [27].

However, recently, some authors used the improved Rosen-Morse potential and improved Tietz potential to represent the internal vibrations of diatomic molecules, including NO [28-30], N2 [31], and CO [32-34], and some triatomic molecules  [35-37], and successfully predicted the vibrational partition functions and important thermodynamic properties for some pure substances. 

In the present work, we study the solutions of the Feinberg-Horodecki equation and extend the subject of coherent states to the space-like coherent states for the temporal counterpart of the Kratzer plus screened Coulomb potential. The motivation of the present work is to obtain the eigen-solution of the Feinberg-Horodecki equation with a  time-dependent modified Kratzer plus screened Coulomb potential by means of the NU method. The rest of this work is organized as follows: the NU method is briefly introduced in Section 2. The approximate solution of the FH equation for the time-dependent general form of Kratzer potential plus screened Coulomb potential is solved to obtain its quantized momentum states and eigenfunctions in Section 3. 
We generate the solutions of a few special potentials mainly found from our general form solution in section 4. Finally we present our discussions and conclusions.

\section{Feinberg-Horodecki equation with time-dependent combined Kratzer plus screened Coulomb potential}
\noindent It is necessary to state that as the potential  $V=V(t)$ depend on $t$, then
\begin{equation}
E=T+V(r,t), 
\end{equation}
is not a constant of motion, the mechanical energy,
\begin{equation}
\frac{dE}{dt}=\frac{d(\frac{1}{2}mv^2)}{dt}+\frac{dV(r,t)}{dt} = \frac{\partial V}{\partial t}
\end{equation}
Here "Conservation of energy" is a universal principle of physics. If V depends on time, then energy is still conserved and, therefore, the energy must be changing in another part of the system in order to be conserved. On the other hand, if the potential is independent of time, then the energy is a constant of the motion, i.e. energy is conserved. 
\begin{equation}
\frac{dE}{dt}=\frac{d(\frac{1}{2}mv^2)}{dt}+\frac{dV(r)}{dt} = 0
\end{equation}
 A very general principle in modern theoretical physics states that for every symmetry there is a conserved quantity. As examples, translation invariance in time symmetry implies a conservation energy and spatial translation symmetry implies a conservation in momentum and so forth.

The Nikiforov-Uvarov (NU) method [38] is being applied to find the approximate solutions of FH equation for the Kratzer plus screened Coulomb potential then the eigenvalues and eigenfunctions of two special cases are obtained from the results.

The time-dependent Kratzer potential is given by [39-41]
\begin{equation}
V(t)= D- \frac{B}{t} + \frac{q C}{t^2},
\label{eq:KFO}
\end{equation}
where D , B, and C are adjustable real potential parameters, and q is dimensionless parameter. 
Further, the screened Coulomb potential is defined as [26]
\begin{equation}
V(t)= \frac{A e^{-\alpha t}}{t}
\label{eq:coul. pot}
\end{equation}
If equations (\ref{eq:KFO}) and (\ref{eq:coul. pot}) are substituted in FH equation, one obtains
\begin{equation}
\left[-\frac{\hbar^2}{2mc^2}\frac{d^2}{dt^2}+ \left(D- \frac{B}{t} + \frac{q C}{t^2} + \frac{A e^{-\alpha t}}{t}\right)\right]\psi_n(t) = cP_n \psi_n(t),
\end{equation}
where c is the speed of light and $P_n (n=0, 1, 2, ....)$ is the momentum eigenvalues.

Using the Greene-Aldrich approximation [26] defined as
\begin{equation}
\frac{1}{t} \approx \frac{\alpha}{(1-e^{-\alpha t})},
\end{equation}
then, letting $s=e^{-\alpha(t)}$, where s $\epsilon$(0, 1), one gets
\begin{equation}
\frac{d^2\psi_n(s)}{ds^2} + \frac{(1-s)}{s(1-s)}\frac{d\psi_n(s)}{ds}+\frac{-\epsilon_1^2-\epsilon_3s+\epsilon_2s^2}{s^2(1-s)^2}\psi_n(s)=0,
\label{eq:differential eq to solve}
\end{equation}
where the equation is satisfying the asymptotic behaviors where $\psi_n(s = 0) = 0$ and $\psi_n(s = 1) = 0$ and 
\begin{equation}
\epsilon_1^2=\frac{2mc^2}{\hbar^2 \alpha^2}\left(D - B \alpha + q C \alpha^2 - cP_n\right),
\label{eq:gamma1}
\end{equation}
\begin{equation}
\epsilon_2=\frac{ 2mc^2}{\hbar^2 \alpha^2}\left(A\alpha - D +cP_n\right),
\end{equation}
\begin{equation}
\epsilon_3=\frac{2mc^2}{\hbar^2 \alpha^2}\left(B \alpha +A \alpha - 2D +2cP_n\right).
\end{equation}
After comparing equation (\ref{eq:differential eq to solve}) with equation (\ref{eq: NU-equ}), one obtains \\
$\tilde{\tau}(s)=1-s$, $\sigma(s)= s(1-s)$, and $\tilde{\sigma}(s)=-\epsilon_1^2-\epsilon_3s+\epsilon_2 s^2$.\\
When these values are substituted in equation (\ref{eq:pi def.}), we get
\begin{equation}
\pi(s)= -\frac{s}{2}\pm \sqrt{\left(\frac{1}{4}-\epsilon_2-k\right)s^2+ (k+\epsilon_3)s+\epsilon_1^2}.
\label{eq: pi solving}
\end{equation}
As mentioned in the NU method, the discriminant under the square root, in equation (\ref{eq: pi solving}), has to be zero, so that the expression of $\pi (s)$ becomes the square root of a polynomial of the first degree. This condition can be written as
\begin{equation}
\left(\frac{1}{4}-\epsilon_2-k\right)s^2+ (k+\epsilon_3)s+\epsilon_1^2=0.
\end{equation}
After solving this equation, we get
\begin{equation}
s= \frac{-(k+\epsilon_3)\pm\sqrt{(k+\epsilon_3)^2-4\epsilon_1^2(\frac{1}{4}-\epsilon_2-k)}}{2(\frac{1}{4}-\epsilon_2-k)}.
\label{eq: s}
\end{equation}
Then, for our purpose we assume that
\begin{equation}
(k+\epsilon_3)^2-4\epsilon_1^2\left(\frac{1}{4}-\epsilon_2-k\right)=0.
\end{equation}
Arranging this equation and solving it to get an expression for k which is given by the following,
\begin{equation}
k_\pm = -\epsilon_3-2\epsilon_1^2\pm 2\epsilon_1\left(\frac{1}{Q}-\frac{1}{2}\right),
\end{equation}
where the expression between the parentheses is given by
\begin{equation}
\frac{1}{Q}=\frac{1}{2} +\sqrt{\frac{1}{4}+\frac{2mc^2}{\hbar^2 \alpha^2}\left(qC \alpha^2\right)}.
\label{eq:1/Q}
\end{equation}
If we substitute $k_-$ into equation (\ref{eq: pi solving}) we get a possible expression for $\pi(s)$, which is given by 
\begin{equation}
\pi_-(s) =\epsilon_1 -s( \epsilon_1+\frac{1}{Q}),
\label{eq: pi(s) solution}
\end{equation}
this solution satisfy the condition that the derivative of $\tau(s)$ is negative. Therefore, the expression of $\tau(s)$ which satisfies these conditions can be written as
\begin{equation}
\tau(s) =1-s +2\epsilon_1-2s(\epsilon_1+\frac{1}{Q}).
\label{eq: tau(s) result}
\end{equation}
Now, substituting the values of $\tau^{'}_-(s)$, $\sigma^{''}(s)$, $\pi_-(s)$ and $k_-$ into equations (\ref{eq: lambda1}) and (\ref{eq: lambda2}), we obtain
\begin{equation}
\lambda_n= \frac{2mc^2}{\hbar^2 \alpha}\left(-A+B-2qC\alpha\right)+\frac{2\epsilon_1}{Q}-\frac{1}{Q},
\label{eq:lambda_n solution}
\end{equation}
and 
\begin{equation}
\lambda= \lambda_n= n(n+ \frac{2}{Q})+ 2n \epsilon_1.
\label{eq:lambda-solution}
\end{equation}
Now, from equations (\ref{eq:lambda_n solution}) and (\ref{eq:lambda-solution}), we get the eigenvalues of the quantized momentum as
\begin{equation}
P_n= \frac{1}{c}\left[(D-B\alpha+qC\alpha^2)-\frac{\alpha^2 \hbar^2}{2mc^2}\left(\frac{\frac{2mc^2}{\alpha \hbar^2}(A-B+2qC\alpha)+n(n+\frac{2}{Q})+\frac{1}{Q}}{2(n+\frac{1}{Q})}\right)^2\right].
\label{eq:eigenvalues of Pn}
\end{equation}
Obviously, the quantized momentum eigenvalues of the time dependent FH equation are dependent on the values of the potential parameters. The sign of the momentum eigenvalues is dependent on the strength of these parameters and their signs. 

Due to the NU method used in getting the eigenvalues, the polynomial solutions of the hypergeometric function $y_n(s)$ depend on the weight function $\rho(s)$ which can be determined by solving equation (\ref{eq:rho differential}) to get 
\begin{equation}
\rho(s)= s^{2\epsilon_1}(1-s)^{(\frac{2}{Q})-1}.
\label{eq:rho-result}
\end{equation}
Substituting $\rho(s)$ into equation (\ref{eq: yn}), we get an expression for the wave functions as
\begin{equation}
y_n(s)= A_n s^{-2\epsilon_1}(1-s)^{-(\frac{2}{Q}-1)}\frac{d^n}{ds^n}\left[s^{n+2\epsilon_1}(1-s)^{n+\frac{2}{Q}-1}\right],
\label{eq: yn 2}
\end{equation}
where $A_n$ is the normalization constant. Solving equation (\ref{eq: yn 2}) gives the final form of the wave function in terms of the Jacobi polynomial $P_n^{(\alpha, \beta)}$ as follows,
\begin{equation}
y_{n}(s)= A_n n! P_n^{(2\epsilon_1, \frac{2}{Q}-1)}(1-2s).
\label{eq: yn last solution}
\end{equation}
Now, substituting $\pi_-(s)$ and $\sigma(s)$ into equation (\ref{eq:pi mentioned}) and then solving it we obtain
\begin{equation}
\phi_n(s)= s^{\epsilon_1}(1-s)^\frac{1}{Q}.
\label{eq: phi solutin}
\end{equation}
Substituting equations (\ref{eq: yn last solution}) and (\ref{eq: phi solutin}) into equation (\ref{eq: psi}), and using $s=e^{-\alpha(t)}$ one obtains,
\begin{equation}
\psi_n(t)= B_n e^{-\alpha\epsilon_1 t}(1-e^{-\alpha t})^\frac{1}{Q} P_n^{(2\epsilon_1, \frac{2}{Q}-1)}(1-2e^{-\alpha t}),
\end{equation}
where $B_n$ is the normalization constant with $\frac{1}{Q}$ is given in equation (\ref{eq:1/Q}). Obviously, the
above wave function is finite at both $t = 0$ and $t \longrightarrow \infty$.
\section{Special cases}
\noindent 
\subsection{The time-dependent modified Kratzer potential}
\noindent 
To get the modified Kratzer potential from the general form, D, A and $\alpha$ are set to zero, $q=1$, $B=2t_eD_e$ and $C={t_e}^2{D_e}$ and then substituted in (\ref{eq:KFO}) to reduce the general form to the special case [25],
\begin{equation}
V(t)=\frac{-2t_e D_e}{t}+ D_e\left(\frac{t_e}{t}\right)^2,
\end{equation}
where $t_e$ and $D_e$ represent the equilibrium time point and the dissociation energy of the system respectively.
In addition, by substituting the same constants in (\ref{eq:eigenvalues of Pn}) we get the eigenvalues of the time-dependent HF equation with modified Kratzer potential. Our result becomes as follows:
\begin{equation}
P_n= \frac{1}{c}\left[-\frac{2mc^2 }{\hbar^2}\left(\frac{2t_eD_e}{2(n+\frac{1}{Q})}\right)^2\right].
\end{equation}
where
\begin{equation}
\frac{1}{Q}=\frac{1}{2}+\sqrt{\frac{1}{4}+\frac{2mc^2 t_e^2 D_e}{\hbar^2}}.
\end{equation}

On the other hand, to determine the eigenfunctions associated with the modified Kratzer potential, the same parameters were substituted in (\ref{eq:gamma1}) which results in 
\begin{equation}
\psi_n(s)= B_n e^{-\alpha \epsilon_1 t} (1-e^{-\alpha t})^\frac{1}{Q} P_n^{(2\epsilon_1, \frac{2}{Q}-1)}(1-2e^{-\alpha t}),
\end{equation}
where 
\begin{equation}
\epsilon_1=\sqrt{\frac{2mc^2}{\hbar^2 \alpha^2}\left(- 2t_eD_e \alpha + t_e^2D_e \alpha^2 - cP_n\right)},
\end{equation}

\subsection{The time-dependent screened Coulomb potential}
\noindent

 Further by setting the values of B, C, and D to zero and A to -$Ze^2$ we get the screened coulomb potential. And by substituting these values in equation (\ref{eq:eigenvalues of Pn}), it gives the eigenvalues of the FH time dependent equation. These eigenvalues are given by the relation,
\begin{equation}
P_n= \frac{1}{c}\left[-\frac{\alpha^2 \hbar^2}{2mc^2}\left(\frac{\frac{-2mc^2Ze^2}{\alpha \hbar^2}+n(n+2)+1}{2(n+1)}\right)^2\right].
\end{equation}

To determine the eigenfunctions associated with the screened Coulomb potential, the same parameters were substituted in (\ref{eq:gamma1}) which results in 
\begin{equation}
\psi_n(s)= B_n e^{-\alpha \epsilon_1 t}(1-e^{-\alpha t}) P_n^{(2\epsilon_1, 1)}(1-2e^{-\alpha t}),
\end{equation}
where 
\begin{equation}
\epsilon_1=\frac{c}{\hbar \alpha}\sqrt{2mcP_n}.
\end{equation}

Setting $\alpha=0$ in the screened Coulomb potential, one gets the quantized eigenvalues of FH equation with Coulomb potential. Our result for the Coulomb potential is as follow:
\begin{equation}
P_n= \frac{-mc^2z^2e^4}{2\hbar^2(n+1)^2}
\label{eq:coulomb}
\end{equation}

\subsection{Numerical results and discussion}
We compute the momentum eigenvalues of
time dependent Kratzer plus screened Coulomb potential for some diatomic molecules like $CO$, $NO$, $O_2$ and $I_2$. This was done using the spectroscopic parameters displayed in Table \ref{table:paraeter}.

\begin {table}[H]
{\scriptsize
	\caption {Spectroscopic parameters of the various diatomic molecules [43]} \label{tab:Parameters} 
	\vspace*{12pt}
\begin{center}
	\begin{tabular}{c  c   c  c }
		
		\hline
		 \\
		Molecule & $D_e$ (eV) & $t_e$ (time unit) & $\mu$ (a.m.u) \\\\
		\hline\\
		CO & 10.84514471 &  1.1282 & 6.860586000   \\\\
		
		NO & 8.043782568 &  1.1508 & 7.468441000   \\\\
		
		O$_2$  & 5.156658828 &  1.2080 & 7.997457504   \\\\
		
		I$_2$ & 1.581791863 &  2.6620 & 63.45223502  \\
			
		\end {tabular}
		\label{table:paraeter}
	\end{center}
	\vspace*{-2pt}}
\end{table}

Its worth noting that when we solve the FH equation in the absence of the interaction potential, i.e., $V(t)=0$ the quantized momentum eigenvalues are negative. Further, in the presence of Coulomb interaction potential $V(t)=-ze^2/t$, the quantized momentum still stand negative as shown by equation (\ref{eq:coulomb}).
On the other hand, if we let the diatomic molecule interact via the Kratzer plus Screened Coulomb potential this interaction shift the quantized spectrum to the positive region at small values of the screening parameter $\alpha$.
The momentum spectrum for states tends to be continuous spectrum as state n is increasing for the studied diatomic molecules.
The momentum spacing increases with increasing the screening parameter $\alpha$ for all molecules. The momentum spacing between states decreases with increasing n. This momentum spacing difference between states is largest for $CO$, $NO$ and $O_2$, respectively, while smallest for $I_2$ as in Tables \ref{table:Numerical}-\ref{table:Numerical3}. This spacing is almost same for each molecule for different values of screening parameter. For example, for $CO$, $NO$, $O_2$ and $I_2$,  $P_1-P_0= 1.20$, $1.00$, $0.60$ and $0.07eV/c$, respectively. However, in $P_9-P_8= 0.40$, $0.20$, $0.13$ and $0.04$ $eV/c$.

\begin {table}[H]
\centering
{\scriptsize
	\caption {The FH quantized momentum eigenvalues (in units eV/c) for the Kratzer plus screened Coulomb potential for diatomic molecules ($\alpha=0.001$, $D_e$, $t_e$ and $\mu$ as defined in Table \ref{table:paraeter}).} \label{tab:Numerical} 
	\vspace*{12pt}
	\begin{center}
		\begin{tabular}{|p{1cm}|c|c|c|c|}
			\hline
			&\multicolumn{4}{|c|}{$\alpha=0.001$} \\
			\hline \centering
			n &  $CO$ & $NO$ & $O_2$ & $I_2$ \\
			 \hline \centering
			0 & 1.540356974	 & 1.395600000	 & 1.176000000 & 0.386600000\\
			\hline \centering
			1 &2.700454092  &  2.305000000	 &  1.789300000& 0.447000000\\
			\hline \centering
			2 &3.656651169  & 3.040100000 & 2.271200000	 &  0.503000000	 \\
			\hline \centering
			3  & 4.454075791 & 3.642700000	 & 2.656900000	& 0.554900000  \\
			\hline \centering
			4 &5.126028125  &4.142800000  &2.970200000	 & 0.603200000	 \\
			\hline \centering
			5 &5.697514722  &4.562400000   &3.228200000	  & 0.648200000 	 \\
			\hline \centering
			6 &6.187609162  &4.918000000  & 3.443300000	 & 0.690100000	  \\
			\hline \centering
			7  & 6.611064425 & 5.221900000 &3.624400000	 &  0.729300000	 \\
			\hline \centering
			8 &6.979436269  & 5.483600000	 &3.778300000	 & 0.766000000	 \\
			\hline \centering
			9 & 7.301880048	 & 5.710700000  & 3.910300000	 & 0.800400000 	 \\
			\hline 
			
			\end {tabular}

			\label{table:Numerical}
		\end{center}
	\vspace*{-2pt}}
\end{table}

\begin {table}[H]
{\scriptsize
	\caption {The FH quantized momentum eigenvalues (in units eV/c) for the Kratzer plus screened Coulomb potential for diatomic molecules ($\alpha=0.05$, $D_e$, $t_e$ and $\mu$ as defined in table \ref{table:paraeter}).} \label{tab:Numerical2} 
	\vspace*{12pt}
	\begin{center}
		\begin{tabular}{|p{1cm}|c|c|c|c|}
			\hline
			&\multicolumn{4}{|c|}{$\alpha=0.05$} \\
			\hline
			\centering
			n & $CO$ & $NO$ &$O_2$ & $I_2$ \\
			\hline \centering
			0 & 1.629666341	 & 1.477300000  & 1.249400000	 & 0.440300000	 \\
			\hline \centering
			1 &	2.857499074	  &  2.440500000	 &  1.901800000	& 0.509600000	 \\
			\hline \centering
			2 &3.869194581	  & 3.218800000	 & 2.414300000	 &  0.573800000	 \\
			\hline \centering
			3  & 4.712568654	 & 3.856500000 & 2.824000000	& 0.633300000	  \\
			\hline \centering
			4 &5.422905120	  &4.385500000	  &	3.156600000 & 0.688600000	 \\
			\hline \centering
			5 &	6.026700142	  &	 4.829000000	  &	 3.430300000	 &  0.740000000	 \\
			\hline \centering
			6 &	 6.544163497	 &	 5.204500000	 & 3.658100000	 & 0.788000000  \\
			\hline \centering
			7  & 6.990926983	 & 5.525100000	 &	3.849600000	 &  0.832700000	 \\
			\hline \centering
			8 &	7.379234725  & 5.801000000	 &4.012100000	 & 0.874600000		 \\
			\hline \centering
			9 & 7.718787450 & 6.040000000	  & 4.151100000	 & 0.913800000 	 \\
			\hline
			
			\end {tabular}
			\label{table:Numerical2}
		\end{center}
		\vspace*{-2pt}}
\end{table}

\begin {table}[H]
{\scriptsize
	\caption {The FH quantized momentum eigenvalues (in units eV/c) for the Kratzer plus screened Coulomb potential for diatomic molecules ($\alpha=0.1$, $D_e$, $t_e$ and $\mu$ as defined in table \ref{table:paraeter}).} \label{tab:Numerical3} 
	\vspace*{12pt}
	\begin{center}
		\begin{tabular}{|p{1cm}|c|c|c|c|}
			\hline
			&\multicolumn{4}{|c|}{$\alpha=0.1$} \\
			\hline \centering
			n & $CO$ & $NO$ &$O_2$ & $I_2$ \\
			\hline \centering
			0 &  1.720708160	& 1.560500000  & 1.324200000 & 	0.495200000 \\
			\hline \centering
			1 &	3.017005473  & 2.578100000	 &  2.016100000	& 0.573500000 \\
			\hline \centering
			2 &	 4.084136054 & 3.399400000 & 2.558600000&  0.645800000 \\
			\hline \centering
			3  & 4.972738926 & 4.071500000 & 2.991600000& 0.712800000	  \\
			\hline \centering
			4 &	5.720181209  &	4.62800000  &	3.342300000 & 0.774900000	 \\
			\hline \centering
			5 &	6.354523505  &	 5.093800000  &	 3.630000000 &  0.832700000	 \\
			\hline \centering
			6 &	 6.897168854	 &	 5.487200000 & 3.868600000	 & 0.886400000	  \\
			\hline \centering
			7  & 7.364672017 & 5.822200000 &	4.068300000 &  0.936400000	 \\
			\hline \centering
			8 &	7.770000059  & 6.109500000	 &4.237000000	 & 0.983100000		 \\
			\hline \centering
			9 & 8.123426476 & 6.357500000  & 4.380400000 & 1.026600000 	 \\
			\hline
			
			\end {tabular}
			\label{table:Numerical3}
		\end{center}
		\vspace*{-2pt}}
\end{table}

\begin{figure}[H]
	\includegraphics[width=1.05\linewidth]{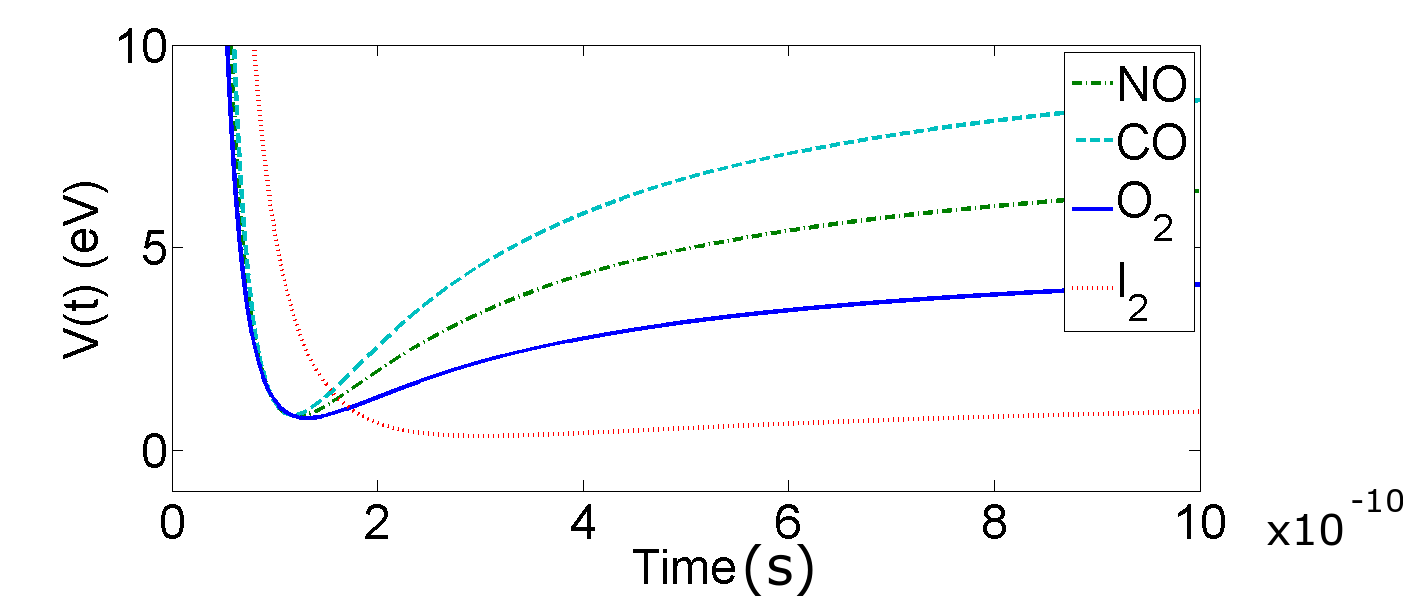}
	\caption[KSC Potential]{Kratzer plus screened Coulomb potential for diatomic molecules.}
	\label{fig:KSC_P_time}
\end{figure}

\begin{figure}[H]
	\includegraphics[width=1.05\linewidth]{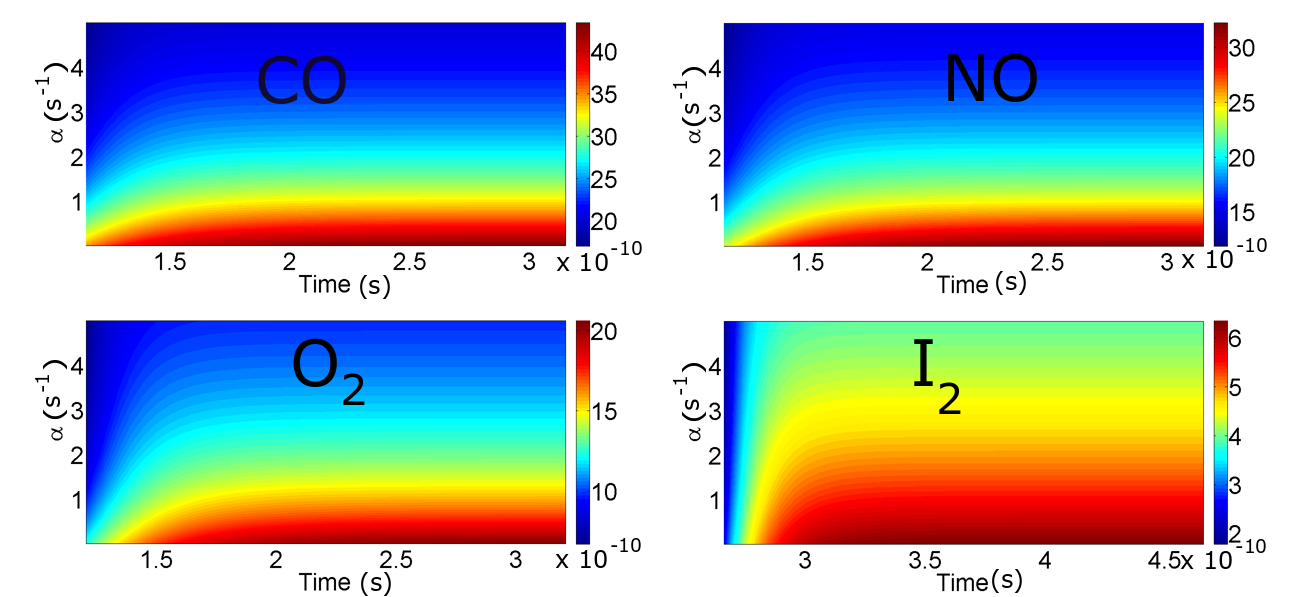}
	\caption[KSC Potential]{Kratzer plus screened Coulomb potential plotted vs both time and screening parameter $\alpha$ for diatomic molecules. Colors represent the value of the potential as illustrated by the color bar.}
	\label{fig:KSCP}
\end{figure}

\begin{figure}[H]
	\includegraphics[width=1.05\linewidth]{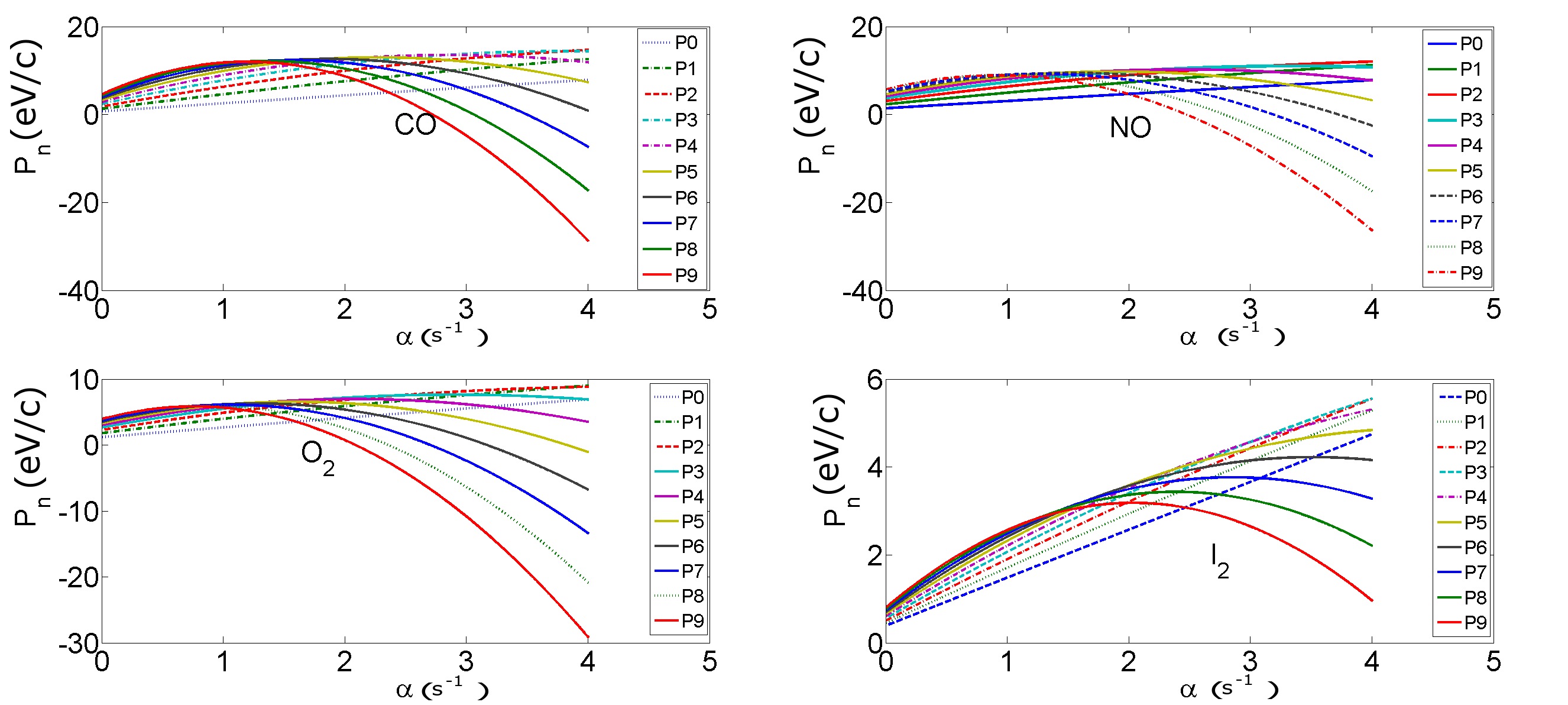}
	\caption[KSC Potential]{FH quantized momentum eigenvalues for the Kratzer plus screened Coulomb potential for diatomic molecules.}
	\label{fig:KSC_Pn}
\end{figure}

 In Figure \ref{fig:KSC_P_time}, we plot the time-dependent Kratzer plus screened Coulomb potential for four different diatomic molecules. The behavior of these diatomic molecules are relatively similar with little difference for $I_2$. In Figure \ref{fig:KSCP}, the Kratzer plus screened Coulomb potential is plotted versus both time and screening parameter  $\alpha$ for the diatomic
molecules. Here colors represent the value of the potential as illustrated by the color bar.
Figure \ref{fig:KSC_Pn},  FH quantized momentum states for the Kratzer plus screened Coulomb potential are plotted versus the screening parameter $\alpha$ for the diatomic molecules (Table \ref{table:paraeter}). We see that the momentum reaches its continuous value when n increases. We sharpen our analysis by taking n=9, we found from Figure \ref{fig:KSC_Pn} that momentum reaches the continuous spectrum when $2<\alpha<3$ for $CO$, $\alpha$ $\approx$ $2$ for $NO$, $1<\alpha <2$ for $O_2$ and $\alpha >4$ for $I_2$.

\begin{figure}[H]
	\includegraphics[width=1.05\linewidth]{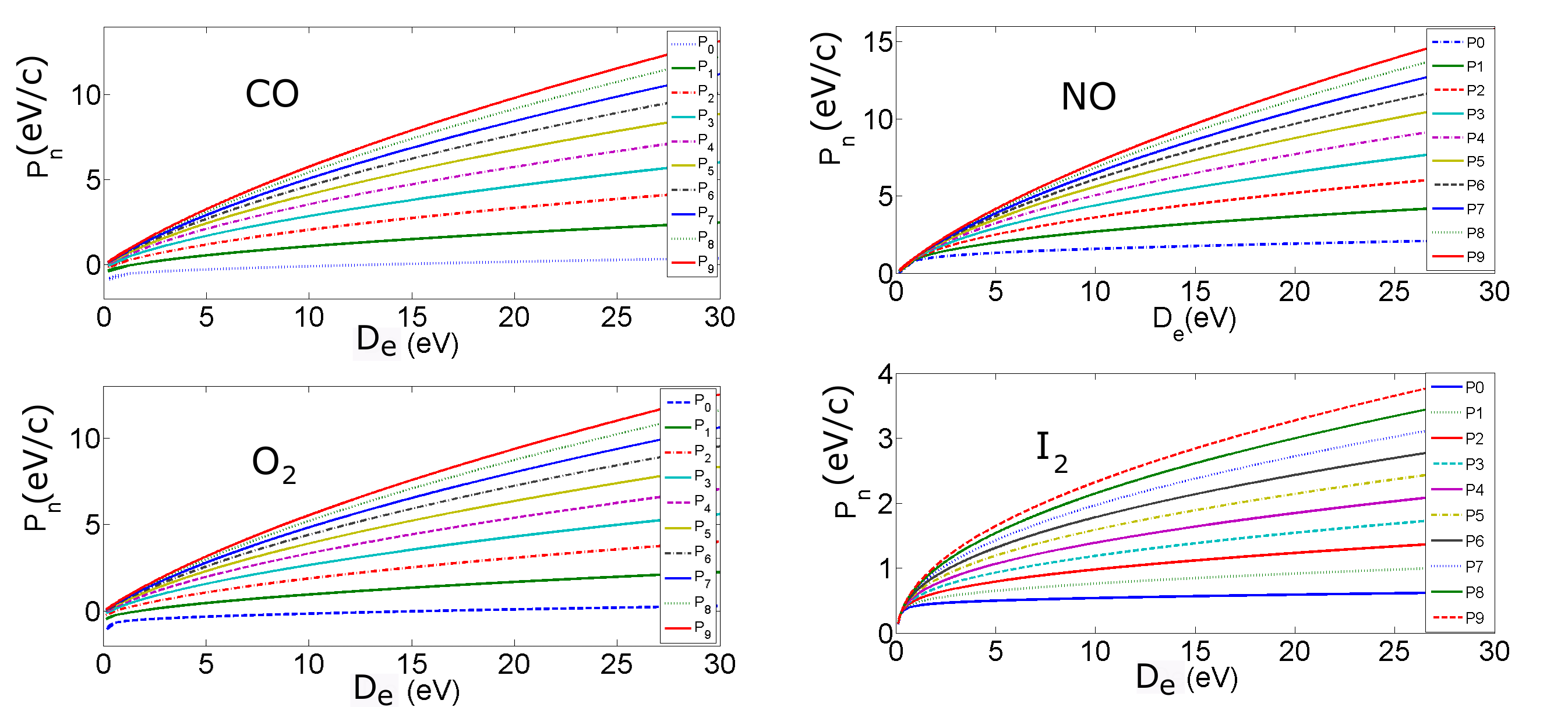}
	\caption[KSC Potential]{FH quantized momentum eigenvalues for the Kratzer plus screened Coulomb potential for diatomic molecules. The graphs are plotted vs De for $\alpha =0.005$.}
	\label{fig:KSCP_De}
\end{figure}

\begin{figure}[H]
	\includegraphics[width=1.05\linewidth]{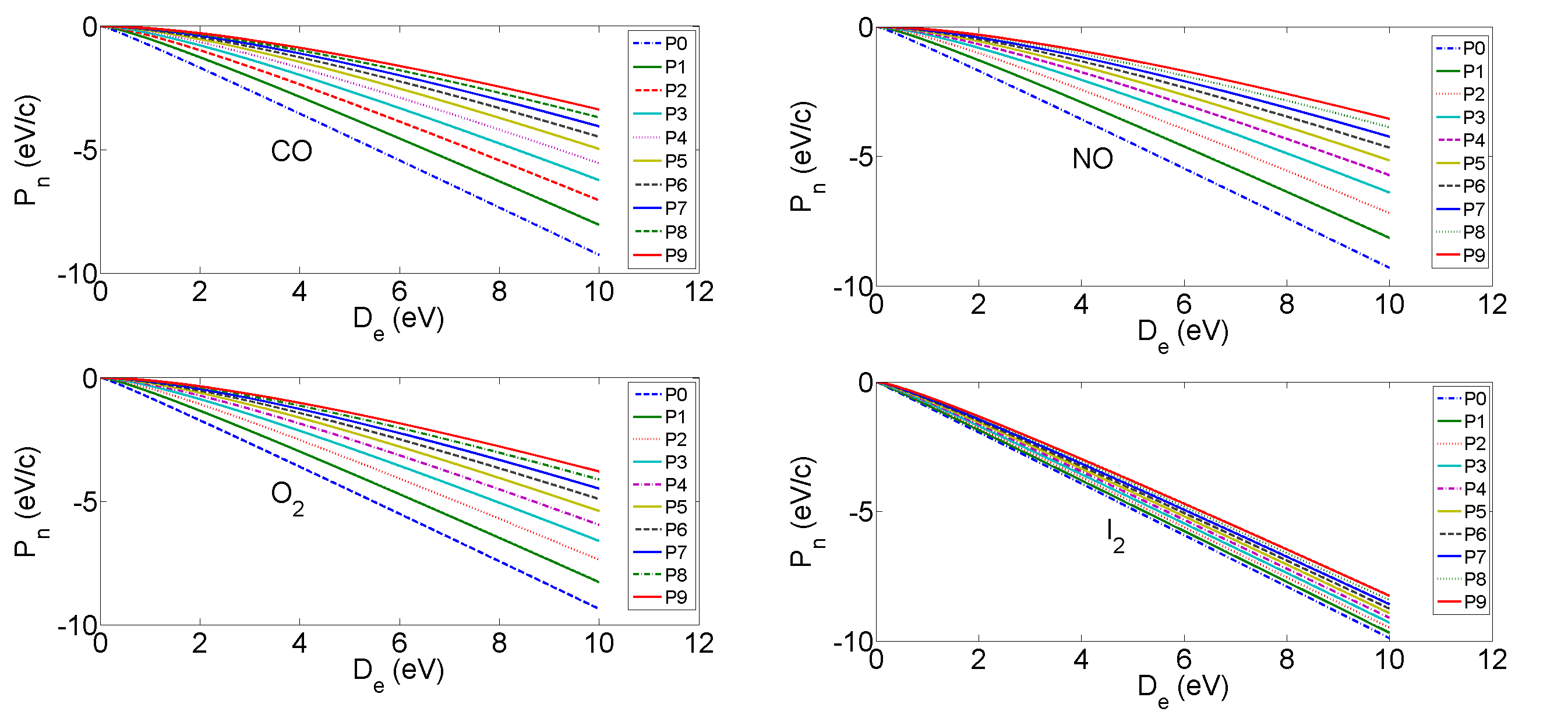}
	\caption[KSC Potential]{FH quantized momentum eigenvalues for the modified Kratzer potential for diatomic molecules.}
	\label{fig:K_PnDe}
\end{figure}

\begin{figure}[H]
	\includegraphics[width=1.05\linewidth]{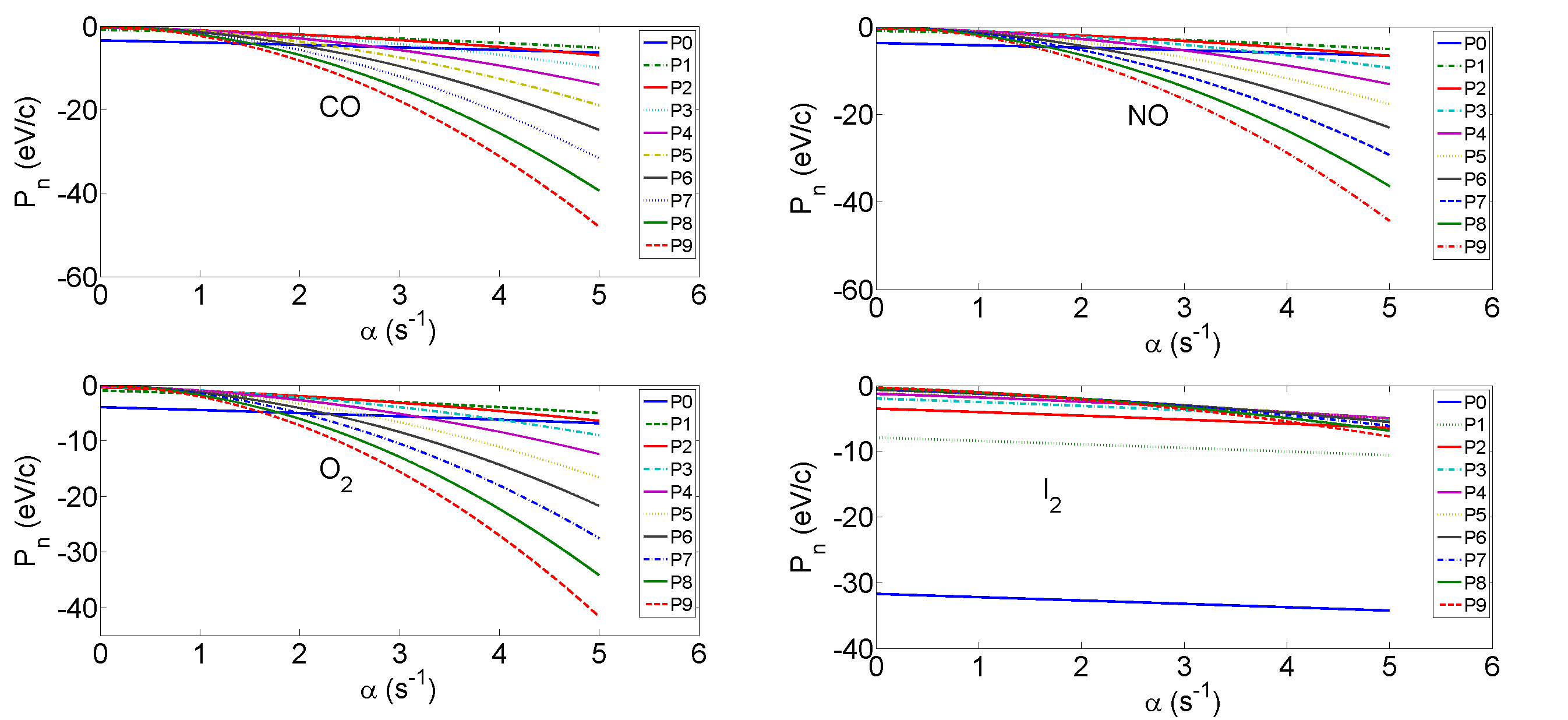}
	\caption[KSC Potential]{FH quantized momentum eigenvalues of the screened Coulomb potential for diatomic molecules.}
	\label{fig:Coul_Pn_alpha}
\end{figure}

In Figure \ref{fig:KSCP_De}, the FH quantized momentum eigenvalues for the Kratzer plus screened Coulomb potential are plotted versus the potential strength $D_e$ for $\alpha=0.005$ $(1/s)$, the momentum increases above $10$ $eV/c$ for $CO$, $NO$ and $O_2$ whereas for $I_2$ it reaches to $4$ $eV/c$. Figure \ref{fig:K_PnDe} shows the FH quantized momentum spectrum versus the potential parameter $D_e$ in the negative region for modified Kratzer potential for diatomic molecules. Finally, in Figure \ref{fig:Coul_Pn_alpha}, the FH quantized momentum eigenvalues of screened Coulomb potential versus the screening parameter for diatomic molecules reach the continuous region for small value of $\alpha <1$.

\section{Conclusions}
\noindent We solved the Feinberg-Horodecki (FH) equation for the time-dependent general form of Kratzer potential via Nikiforov-Uvarov (NU) method. We got the approximate quantized momentum eigenvalues solution of the FH equation. It is therefore, worth mentioning that the method is elegant and powerful. Our results can be applied in biophysics and other branches of physics.
In this paper, we have applied our result for the modified Kratzer and screened coulomb potentials, as special cases of the used  potential, for quantized momentum eigenvalues. 

\section*{References}
\noindent [1] Park T J 2002 \textit{Bull. Korean Chem. Soc.} \textbf{23} 1733

\noindent [2] Vorobeichik I, Lefebvre R, and Moiseyev N 1998 \textit{EPL} \textbf{41} 111

\noindent [3] Shen J Q 2003 arXiv:0310179 [quant-ph]

\noindent [4] Feng M 2001 \textit{Phys. Rev. A} \textbf{64} 034101 

\noindent [5] Horodecki R 1988 \textit{Il Nuovo Cimento B} \textbf{102} 27

\noindent [6] Feinberg G 1967 \textit{Phys. Rev.} \textbf{159} 1089

\noindent [7] Molski M 2006 \textit{The Eur. Phys. J. D.} \textbf{40} 411

\noindent [8] Molski M 2010 \textit{Biosystems} \textbf{100} 47

\noindent [9] Witten E 1981 \textit{Nuc. Phys. B} \textbf{188} 513

\noindent [10] Molski M 1988 \textit{Phys. J. B: At. Mol. Opt. Phys.} \textbf{21} 3449 

\noindent [11] Recami E and Mignani R 1974 \textit{Riv. Nuovo Cim.} \textbf{4} 209 

\noindent [12] Recami E 1986 \textit{Riv. Nuovo Cim.} \textbf{9} 1 

\noindent [13] Molski M 1999 \textit{Europhys. Lett.} \textbf{48} 115

\noindent [14] Hamzavi M, Ikhdair S M and Amirfakhrian M 2013 \textit{Theor. and App. Phys. J.} \textbf{7} 40

\noindent [15] Eshghi M, Sever R, and Ikhdair S M 2016 \textit{Eur. Phys. J. Plus} \textbf{131} 223

\noindent [16] Berkdemir C, Berkdemir A, and Han J 2006 \textit{Chem. Phys. Lett.} \textbf{417} 326

\noindent [17] Sadeghi J 2007 \textit{Acta Phys. Polon.} \textbf{112} 23

\noindent [18] Sever R and Tezcan C 2008 \textit{Int. J. Mod. Phys. E} \textbf{17} 1327

\noindent [19] Kandirmaz N 2018 \textit{Math. Phys. J.} \textbf{59} 063510

\noindent [20] Cheng Y F and Dai T Q 2007 \textit{Phys. Scr.} \textbf{75} 274

\noindent [21] Hassanabadi H, Rahimov H, and Zarrinkamar S 2011 \textit{Advances in High Energy Physics} \textbf{2011}

\noindent [22] Yan-Fu C and Tong-Qing D 2007 \textit{Commun. Theor. Phys.} \textbf{48} 431

\noindent [23] Ghodgaonkar A and Ramani K 1981 \textit{J. Chem. Soc. Faraday Trans.} \textbf{77} 209

\noindent [24] Khordad R 2013 \textit{Indian J. Phys.} \textbf{87} 623

\noindent [25] Babaei-Brojeny A A and Mokari M 2011 \textit{Phys. Scr.} \textbf{84} 045003

\noindent [26] Edet C, Okorie U, Ngiangia A, and Ikot A 2019 \textit{Indian J.Phys.} 1

\noindent [27] Okorie U, Edet C, Ikot A, Rampho G, and Sever R 2020 \textit{Indian J. Phys}. 1

\noindent [28]  Jia CS, Wang CW, Zhang LH, Peng XL, Tang HM, Zeng R 2018 \textit{Chem. Eng. Sci.} \textbf{183} 26

\noindent [29] Peng XL, Jiang R, Jia CS, Zhang LH, Zhao YL 2018 \textit{Chem. Eng. Sci.} \textbf{190} 122

\noindent [30] Jia CS, Zeng R, Peng XL, Zhang LH, Zhao YL 2018 \textit{Chem. Eng. Sci.} \textbf{190} 1

\noindent [31] Jia CS, Zhang LH, Peng XL, Luo JX, Zhao YL, Liu JY, Guo JJ, Tang LD 2019 \textit{Chem. Eng. Sci.} \textbf{202} 70

\noindent [32] Jia CS, Wang CW, Zhang LH, Peng XL, Zeng R, You XT 2017 \textit{Chem. Phys. Lett.} \textbf{676} 150

\noindent [33] Jia CS, Wang CW, Zhang LH, Peng XL, Tang HM, Liu JY, Xiong Y, Zeng R 2018 \textit{Chem. Phys. Lett.} \textbf{692} 57

\noindent [34] Jiang R, Jia CS, Wang YQ, Peng XL, Zhang LH 2019 \textit{Chem. Phys. Lett.} \textbf{715} 186

\noindent [35] Chen XY, Li J, Jia CS 2019 \textit{ACS Omega}  \textbf{4} 16121

\noindent [36] Wang J, Jia CS, Li CJ, Peng XL, Zhang LH, Liu JY 2019 \textit{ACS Omega} \textbf{4} 19193

\noindent [37] Jia CS, Wang YT, Wei LS, Wang CW, Peng XL, Zhang LH 2019 \textit{ACS Omega} \textbf{4} 20000

\noindent [38] Nikiforov A F and Uvarov V B 1988 \textit{Springer} \textbf{205} (in Russian)

\noindent [39] Kratzer A 1920 \textit{Zeitschrift fur Physik A} \textbf{3} 289 (in Deutsch)

\noindent [40] Molski M 2007 arXiv:0706.3851

\noindent [41] Ikot A N, Okorie U, Ngiagian A Th, Onate C A, Edet C O, Akpan I O and Amadi P O 2020 \textit{Ecl{\'e}tica Qu{\'\i}mica J}. \textbf{45} 65

\section*{}


\appendix
\section*{Appendix: Review to Nikiforov-Uvarov method}

\noindent Nikiforov-Uvarov (NU) [38] method is usually used to reduce the second-order differential equation into a general form of a hypergeometric type. In that sense, any second order differential equation can be transformed, using a suitable coordinate transformation s=s(r), into the form:
\begin{equation}
\psi_n^{''}(s)+ \frac{\tilde{\tau}(s)}{\sigma(s)} \psi_n^{'}(s)+ \frac{\tilde{\sigma}(s)}{\sigma^2(s)} \psi_n(s)=0,
\label{eq: NU-equ}
\end{equation}
where $\sigma (s)$ and $\tilde{\sigma}(s)$ are polynomials, at most second-degree, and $\tilde{\tau}(s)$ is a first-degree polynomial. To solve equation (\ref{eq: NU-equ}), the following wave function can be proposed,
\begin{equation}
\psi_n(s)= \phi_n(s) y_n(s),
\label{eq: psi}
\end{equation}
which transforms equation (\ref{eq: NU-equ}) into the following hypergeometric form
\begin{equation}
\sigma(s) y_n^{''}(s) + \tau(s) y_n^{'}(s) + \lambda y_n(s)= 0,
\label{eq:hypergeometric}
\end{equation}
where \begin{equation}
\sigma(s)= \pi(s) \frac{\phi_n(s)}{{\phi_n}^{'}(s)},
\label{eq:pi mentioned}
\end{equation}
\begin{equation}
\tau(s)= \tilde{\tau}(s) + 2\pi(s),      
\end{equation}
$\lambda$ in equation (\ref{eq:hypergeometric}) is a parameter defined as,
\begin{equation}
\lambda = \lambda_n = -n \tau^{'}(s) - \frac{n(n-1)}{2} \sigma^{''}(s),
\label{eq: lambda1}
\end{equation}
where n=0, 1, 2, ......, and $\tau(s)$ is a polynomial with a negative first derivative to produce an appropriate solution for the hypergeometric equation. 
The first part of the proposed wave function is a solution of the equation,
\begin{equation}
\sigma (s) \phi^{'}(s) - \pi(s) \phi(s) =0.
\end{equation}
Whereas, the second part of the wave function (\ref{eq: psi}) represents a hypergeometric form [24] which can be obtained using Rodrigues formula
\begin{equation}
y_n(s)= \frac{A_n}{\rho(s)} \frac{d^n}{ds^n} [\sigma^n(s) \rho (s)],
\label{eq: yn}
\end{equation}
where $B_n$ is a normalization constant and $\rho(s)$ is a weight function that can be calculated from the relation,
\begin{equation}
\sigma (s) \rho^{'}(s) +[\sigma(s) - \tau(s)] \rho(s) =0.
\label{eq:rho differential}
\end{equation}
$\pi(s)$ in equation (\ref{eq:pi mentioned}) is defined as 
\begin{equation}
\pi(s)= \frac{\sigma^{'}-\tilde{\tau}}{2}\pm \sqrt{(\frac{\sigma^{'}-\tilde{\tau}}{2})^2-\tilde{\sigma}+ k\sigma},
\label{eq:pi def.}
\end{equation}
and $\lambda$ in equation (\ref{eq:hypergeometric}) is defined as,
\begin{equation}
\lambda = \lambda_n= k+ \pi^{'}(s),
\label{eq: lambda2}
\end{equation}
where $\pi(s)$ is a polynomial which depends on the transformation function s(r) and k should be determined to calculate $\pi(s)$, for which the discriminant under the square root in equation (\ref{eq:pi def.}) is set to zero, in order to let $\pi(s)$ to be a first order polynomial. Finally the eigenvalues can be found by solving equations (\ref{eq: lambda1}) and (\ref{eq: lambda2}).

\end{document}